\documentclass[3p,twocolumn]{elsarticle}

\usepackage{graphics}
\usepackage{graphicx}
\usepackage{epsfig}

\usepackage{amssymb}

\journal{Nuclear Instruments and Methods}

\begin{document}

\begin{frontmatter}

\title{Study of a Large NaI(Tl) Crystal}

\author[a]{A. Aguilar-Arevalo}
\author[b]{M. Aoki}
\author[c]{M. Blecher}
\author[d]{D.A. Bryman}
\author[a]{L. Doria\corref{cor1}}
\cortext[cor1]{Corresponding author. luca@triumf.ca (L. Doria)}
\author[a]{P. Gumplinger}
\author[e]{A. Hussein}
\author[b]{N. Ito}
\author[f]{S. Kettell}
\author[a]{L. Kurchaninov}
\author[f]{L. Littenberg}
\author[d]{C. Malbrunot}
\author[a]{G.M. Marshall}
\author[a]{T. Numao\corref{cor2}}
\cortext[cor2]{Corresponding author. toshio@triumf.ca (T. Numao)}
\author[a]{R. Poutissou}
\author[a]{A. Sher}
\author[b]{K. Yamada}

\address[a]{TRIUMF,4004 Wesbrook Mall, Vancouver, B.C. V6T 2A3, Canada}
\address[b]{Physics Department, Osaka University, Toyonaka, Osaka, 560-0043, Japan}
\address[c]{Physics Department, Virginia Tech., Blacksburg, VA 24061, USA}
\address[d]{Department of Physics and Astronomy, University of British Columbia, Vancouver, B.C. V6T 1Z1, Canada}
\address[e]{University of Northern British Columbia, Prince George, B.C. V2N 4Z9, Canada}
\address[f]{Brookhaven National Laboratory, Upton, NY 11973-5000, USA}

\begin{abstract}
Using a narrow band positron beam, the response of a large high-resolution NaI(Tl) crystal 
to an incident positron beam was measured. 
It was found that nuclear interactions cause the appearance of additional peaks in the low 
energy tail of the deposited energy spectrum.
\end{abstract}

\begin{keyword}
Calorimeter \sep Scintillation detectors \sep Photonuclear reactions



\end{keyword}

\end{frontmatter}



\section{Motivation}
The PIENU experiment at TRIUMF \cite{pienu} is aiming at a measurement of the branching ratio
$R=\Gamma (\pi\rightarrow e\nu + \pi\rightarrow e\nu\gamma)/ \Gamma (\pi\rightarrow \mu\nu + \pi\rightarrow \mu\nu\gamma)$
with precision $<$0.1\%.
The principal instrument used to measure  positron energies
from  $\pi^{+} \rightarrow e^{+}\nu$ decays ($E_{e^{+}}=70$~MeV)
and $\pi^{+} \rightarrow \mu^{+} \nu$ followed by $\mu^{+} \rightarrow e^{+} \nu \overline\nu$ 
decays ($E_{e^{+}}=0-53$~MeV) is a large single crystal NaI(Tl) detector \cite{bina}. 
Detailed knowledge of the crystal response is essential to reaching high precision,
especially for determining the low energy tail response below 60~MeV \cite{triumf}. In the following, results 
of measurements of the response of the NaI(Tl) crystal to mono-energetic positron beams are 
presented along with Monte Carlo (MC) simulations including photonuclear reactions.  

\section{Experiment Setup}
\begin{figure}[t]
\resizebox{\columnwidth}{!}{\includegraphics{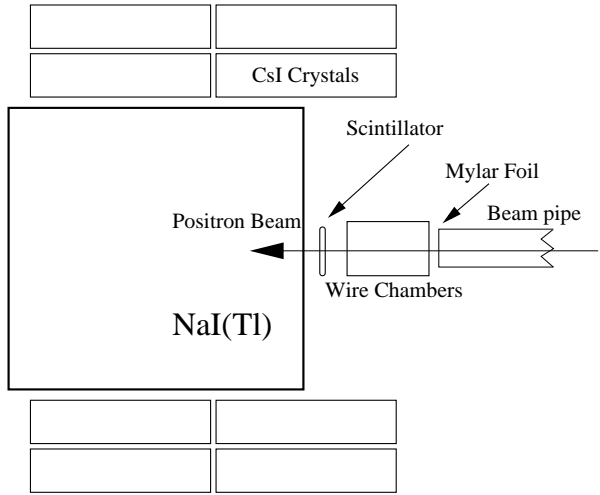}}
\caption{Schematic description of the experimental setup (not to scale).
The beam comes from the right and impinges on the NaI(Tl) crystal surrounded by two rings of 97 CsI crystals.
In front of the NaI(Tl), there are 6 planes of wire chambers and a plastic scintillator.}
\label{setup}
\end{figure}

The 48~cm diameter, 48~cm long NaI(Tl) crystal \cite{bina} under study was surrounded by
two adjacent rings of 97 pure CsI crystals \cite{BNLE787}. 
Each ring was comprised of two layers of 8.5~cm thick, 25~cm long crystals. Positrons
from  the M13 beamline at TRIUMF \cite{m13} were injected into the NaI(Tl) crystal to study its response.
The positrons were produced by 500~MeV protons from the TRIUMF cyclotron striking
a 1~cm thick beryllium target. After defining the beam momentum at the first focus,
the M13 beam line is equipped with two more dipole magnets and two foci with slits before the final focus at the detector.
The vacuum window was a 0.13~mm thick, 15~cm diameter Mylar foil. With this geometry, slit scattering and the 
effect of the vacuum window were expected to have negligible effect on the low energy tail.
The incoming beam was measured with a telescope (see fig.~\ref{setup}) consisting of 6 planes of wire chambers
arranged in the orientation of X-U-V-X-U-V, where U(V) was at $60^{\circ}$($-60^{\circ}$) to the vertical direction, 
a plastic scintillator (5$\times$5~cm$^2$ area, 3.2 mm thickness), and the NaI(Tl) calorimeter.  
The beam momentum width and horizontal (vertical) size and divergence were 1.5\% in FWHM, 2cm (1cm) and $\pm$50mrad ($\pm$90mrad), 
respectively. The beam composition was 63\% $\pi^{+}$, 11\% $\mu^{+}$ and 26\% $e^{+}$.

\section{Measurement and Results}
A 70~MeV/c positron beam was injected into the center of the NaI(Tl) crystal. 
The beam timing with respect to the 23~MHz cyclotron radio frequency provided particle 
identification based on time-of-flight (TOF) together with the energy loss in the beam scintillator,
allowing selection of positrons for studying the crystal response function. 
Events due to positrons from decays of muons previously stopped in the NaI(Tl) 
crystal were suppressed by requiring wire chamber hits, and using TOF and pileup cuts.
Pion and muon contamination was reduced in the data to the 0.08\% level.

The CsI crystals were used in veto mode to select events without shower leakage from the NaI(Tl)
as well as for tagging events with delayed particle emission.
Leakage from the NaI(Tl)'s downstream face was not detected but minimized by the 19 radiation length
thickness of the crystal.

\begin{figure}[t]
\resizebox{\columnwidth}{!}{\includegraphics{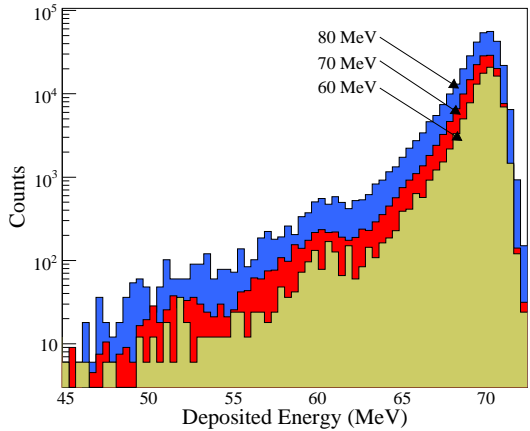}}
\caption{Normalized NaI(Tl) energy spectra for incident positron beam momenta 60, 70, and 80~MeV/c.
The spectra were shifted and aligned to the peak at 70~MeV/c. Histograms are scaled differently
for easier comparison. 
}
\label{momscan}
\end{figure}

The resulting positron energy spectrum is shown in fig.~\ref{momscan} (dark shaded histogram).
The main peak at 70~MeV has an asymmetrical shape due primarily to shower leakage with a width of 2.7\% (FWHM).
Subtracting the calculated beam momentum width in quadrature gave a NaI(Tl) crystal resolution of approximately 2.2\% (FWHM).
Besides the main peak at 70~MeV, there are two additional structures at 62 and 54~MeV.

Studies were made to determine whether the additional peaks had either instrumental or physical origin.
Using different settings of the momentum-defining and collimating slits, which enhanced or suppressed slit scattering,
no effect on the positron energy spectrum was found including the relative intensity of the peaks. 
Also, different tunes of the beamline ({\em e.g.} different focusing) did not change the measured energy spectrum.
The beam momentum was varied in order to observe the corresponding position of the peaks. 
Fig.~\ref{momscan} also shows the spectra for 60 and 80~MeV/c beam momenta shifted and plotted on top of the 
reference histogram at the nominal momentum of 70~MeV/c.
Signals from the CsI crystals were used to suppress the low energy tail due to shower leakage to
enhance the second and third peaks. For all three beam momenta, the relative positions of the low energy peaks remained unchanged. 
The beam position dependence of the NaI(Tl) spectrum was also tested using wire chamber information, without finding any effect.
Based on these tests, it is unlikely that there is an influence of the beam settings in the 
appearance of the additional structures in the energy spectrum.

\begin{figure}[t]
\resizebox{\columnwidth}{!}{\includegraphics{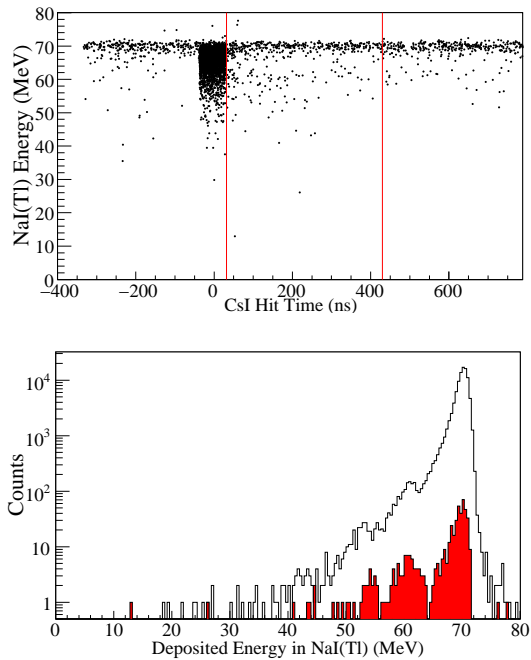}}
\caption{(Top) Deposited energy  versus CsI hit timing.
(Bottom) The shaded histogram represents events selected by the timing cut 
(between the lines) shown on the top figure.}
\label{timing}
\end{figure}

In fig.~\ref{timing} (top), the deposited energy in the NaI(Tl) crystal is shown as a function of the CsI hit time.
The horizontal band at the beam energy corresponds to accidental events,
while the coincident ones from shower leakage are concentrated around 0~ns.
There are delayed events in the low energy region that correspond to the second and third peaks.
If delayed events between the vertical lines are selected, the shaded
spectrum in fig.~\ref{timing} (bottom) is obtained. The first peak (at approximately 70 MeV) 
is consistent with accidental coincidences.
The second and the third peaks were enhanced after the delayed coincidence requirement. 
These results are consistent with the hypothesis of neutrons escaping the NaI(Tl) and giving a delayed 
signal in the CsI. Moreover, the energy deficits of the second and third peaks are consistent with the
separation energies for one ($E_{n}=9.14$ MeV) and two neutrons ($E_{2n}=16.3$ MeV) emitted 
from $^{127}$I. Since only the first hit is plotted in 
fig.~\ref{timing}, the observed secondary peaks are not due to the slow component of the CsI pulse.

The yield of the second peak is consistent with the 30\% solid angle and estimated 10\% detection efficiency
of the CsI calorimeter for  neutron capture. A delay of 100~ns is also consistent with the TOF of $<1$~MeV neutrons.

To estimate the number of neutrons involved in the second and third peaks, two Gaussian functions  
on a background with an exponential shape were fitted to both histograms in fig.~\ref{timing} (bottom).
The ratio $N_{2}$ ($N_{3}$) of the number of events in the second (third) peak before and after the delayed
coincidence requirement is proportional to the product of the neutron detection efficiency and the number 
of neutrons involved, $n_{2}$ ($n_{3}$).
The quantity $N=N_{3}/N_{2} = n_{3}/n_{2}$ indicates the ratio of the neutron multiplicities, which was found
to be $N=2.1 \pm 0.2$. Assuming that one neutron is involved in the second peak, 
this result suggests that the third peak arises when two neutrons escape from the crystal. 

The previous branching ratio experiment \cite{triumf} was not able to 
detect these peaks because of the poorer energy  resolution of the NaI(Tl) crystal ($3-4$\% FWHM) employed.

\begin{figure}[t]
\resizebox{\columnwidth}{!}{\includegraphics{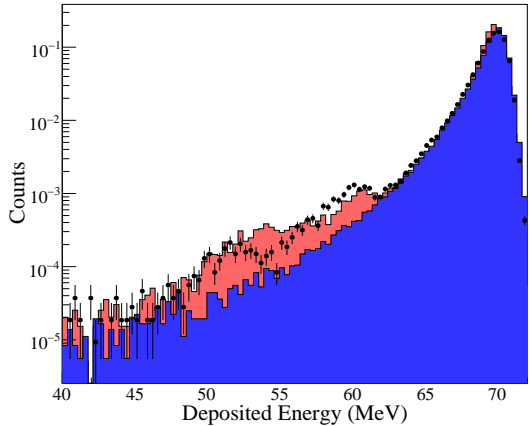}}
\caption{Comparison between data (filled circles with error bars) and simulation. The simulation
was performed with (light shaded) and without (dark shaded) hadronic reaction contributions.
The histograms are normalized to the same area.}
\label{mc}
\end{figure}

\begin{figure}[t]
\resizebox{\columnwidth}{!}{\includegraphics{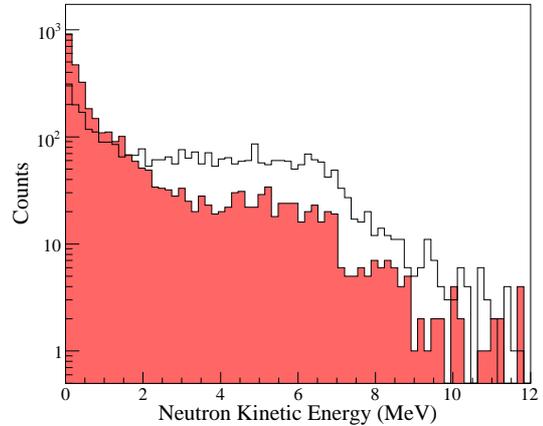}}
\caption{Simulation of the kinetic energy of the neutrons produced in (white histogram) and those that escaped 
from (shaded histogram) the NaI(Tl) crystal.}
\label{nescaped}
\end{figure}

\section{Simulation}
A MC simulation was developed, including all physics
effects available in the GEANT4 package \cite{geant4,geant4-2}. In particular, 
photonuclear reactions with neutron(s) emission, scattering and absorption
were taken into account using the QGSP\_BERT physics processes list.
In fig.~\ref{mc}, the spectra obtained with the same detector setting with a monochromatic beam is shown. 
If in the simulation only electromagnetic interactions were considered 
(dark shaded histogram), the low energy tail shows no structure. 
If hadronic interactions were included, additional structures appear (light shaded histogram), 
which are similar to those observed in the data (filled circles).
A closer look at the simulated data shows that photonuclear reactions followed by neutron escape
from the crystal are indeed responsible for the additional peak structures.
Positrons entering the crystal produce an electromagnetic shower. One or more photons of the shower
can be captured by $^{127}$I nuclei. 
In the MC simulation, nuclear photoabsorption is generally followed by emission of neutrons (94\%), protons (4\%) or 
$\alpha$-particles (2\%). 
The kinetic energy and the separation energy of the neutron are not observed by the NaI(Tl) crystal if the neutron escapes.
The second peak in the deposited energy spectrum starts at $E_{1n}$ below the beam energy, where this
reaction channel opens. 
According to the MC, the origin of the third peak in the spectrum is due to emission and escape of two neutrons.
The neutrons can come from a single nucleus or from two separate ones (due to more photo-absorptions in the same shower). 
Both cases contribute to the third peak which starts at an energy consistent with either the energy 
threshold of two neutron emission or twice the single separation energy $E_{1n}$.

The distribution of the kinetic energy for escaping neutrons is shown in fig.~\ref{nescaped}.
The dashed histogram represents the kinetic energy of the neutrons after nuclear emission, while
the shaded histogram shows the kinetic energy after escape from the NaI(Tl) crystal. 
The difference between the two spectra is due to elastic and inelastic scattering reactions in the NaI(Tl) crystal.

\begin{figure}[!t]
\resizebox{\columnwidth}{!}{\includegraphics{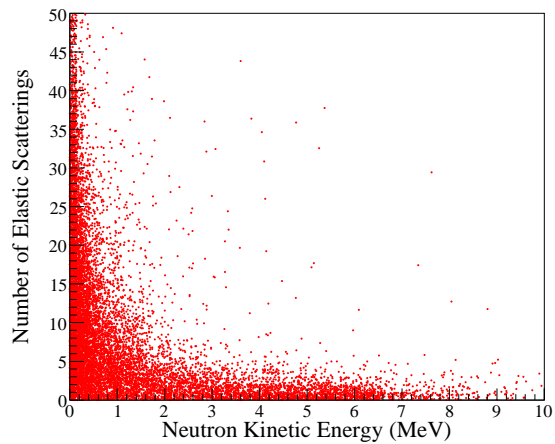}}
\caption{Simulation of the number of elastic scatterings as a function of the kinetic energy of the neutrons after escaping the nucleus.}
\label{elastic}
\end{figure}

In fig. \ref{elastic}, the correlation between the number of elastic scatterings and the neutron kinetic energy at production is shown.
Figs.~\ref{nescaped} and \ref{elastic} suggest that, although the primary source of the second and third peaks 
is low energy neutron emission from photonuclear reactions, 
many neutron elastic scatterings significantly lower the escaping neutron
kinetic energy, returning ``lost'' energy to the NaI(Tl) crystal. 

The agreement between simulation and experiment is not perfect. Given the high number of interactions
which a neutron can experience in a large crystal, a small error in the models for elastic and inelastic
scattering can be amplified. Moreover, in GEANT4 photonuclear reactions are parameterized on a limited data set
of nuclides.

\section{Conclusions}
The response of a large NaI(Tl) crystal to a positron beam of 70~MeV/c was investigated in
preparation for the PIENU experiment at TRIUMF.
Low energy structures were observed in the energy spectrum and the mechanism
for their origin was found to be consistent with neutron emission due to photo-absorption
followed by neutron escape from the crystal.

\section*{Acknowledgments}
We wish to thank M.~Kovash (University of Kentucky) for providing us with his $\gamma$-ray spectrum measured with a similar
NaI(Tl) crystal, A.~Sandorfi (Brookhaven National Laboratory) for useful comments and for arranging 
the loan of the NaI(Tl) crystal, and S.~Chan, C.~Lim and N.~Khan for the engineering and installation work of the detector. 
We are also grateful to Brookhaven National Laboratory for providing the NaI(Tl) and CsI crystals.
This work was supported by the Natural Science and Engineering Council (NSERC)
and the National Research Council of Canada through its contribution to TRIUMF. 
One of the authors (MB) has been supported by US National
Science Foundation grant Phys-0553611.

\end{document}